\documentclass[prb,aps,onecolumn,showpacs,floats,a4paper,times,12pt]{revtex4}
\usepackage{epsfig,bm,epsf,graphics}
\begin{document}
\draft
\title{CROSSOVER FROM FIRST TO SECOND-ORDER TRANSITION IN FRUSTRATED ISING ANTIFERROMAGNETIC FILMS}
\author{X. T. Pham Phu$^{a}$, V. Thanh Ngo$^{b}$, and H. T. Diep$^{a}$\footnote{ Corresponding author, E-mail:diep@u-cergy.fr }}
\address{$^a$ Laboratoire de Physique Th\'eorique et Mod\'elisation,
Universit\'e de Cergy-Pontoise, CNRS, UMR 8089\\
2, Avenue Adolphe Chauvin, 95302 Cergy-Pontoise, France\\
$^b$ Institute of Physics, VAST, P.O. Box 429,   Bo Ho, Hanoi 10000,
Vietnam}

\begin{abstract}
In the bulk state, the Ising FCC antiferromagnet is fully frustrated and
is known to
have a very strong first-order transition.  In this paper,  we study the nature of this phase transition
in the case of a thin film,
as a function of the film thickness. Using Monte Carlo (MC) simulations,
we show that the transition remains first order down to a thickness of four FCC cells.
It becomes clearly  second order at a thickness of two FCC cells, i.e. four atomic layers.
It is also interesting to note that the presence
of the surface reduces the ground state (GS) degeneracy found in
the bulk.  For the two-cell thickness, the surface magnetization is larger than the interior one.
It  undergoes a second-order phase transition at a temperature $T_C$ while interior spins
become disordered at a lower temperature $T_D$. This loss of order is characterized
by a peak of the interior spins susceptibility and a peak of the specific heat which
 do not depend on the lattice size suggesting that either
it is not a real transition or it is a Kosterlitz-Thouless nature.
The surface transition, on the other hand, is shown to be of second order
with critical exponents deviated  from those of pure 2D Ising universality class.
We also show  results obtained from the Green's function method.  Discussion is given.
\end{abstract}
\maketitle
\section{Introduction}

This paper deals with the question whether or not the phase transition known in
the bulk state changes its nature when the system is made as a thin film.
In a recent work, we have considered the case of a bulk second-order transition. We
have shown that under a thin film shape, i.e. with a finite thickness, the transition
shows effective critical exponents whose values are between 2D and 3D universality classes.\cite{Pham}
If we scale these values with a function of thickness as suggested by Fisher\cite{Fisher} we should find,
as long as the thickness is finite, the 2D universality class.

In this paper, we study the effect of the film thickness in the
case of a bulk first-order transition.  The question to which we would like to answer is whether
or not the first order becomes a second order when reducing the thickness. For that
purpose we consider the face-centered cubic (FCC) Ising antiferromagnet. This system
is fully frustrated with a very  strong first-order transition.

On the one hand, effects of the frustration in spin systems have been extensively
investigated during the last 30 years.  In particular, by exact solutions, we have shown that
frustrated spin systems have rich and interesting properties such as successive phase transitions
with complicated nature, partial disorder, reentrance and disorder lines.\cite{diep91honeyc,diep91Kago}
Frustrated systems still challenge
theoretical and experimental methods. For  recent reviews, the
reader is referred to Ref. \onlinecite{Diep2005}.

On the other hand, physics of surfaces and objects of nanometric
size have also attracted an immense interest.  This is
due to important applications in
industry.\cite{zangwill,bland-heinrich,Diehl} In this
field, research results are often immediately used
for industrial applications, without waiting for a full
theoretical understanding. An example is the so-called giant
magneto-resistance (GMR) used in data storage devices, magnetic sensors,
... \cite{Baibich,Grunberg,Fert,review}  In parallel to these
experimental developments, much theoretical effort has also been devoted to
the search of physical mechanisms lying behind new properties found
in nanometric objects such as ultrathin films, ultrafine particles, quantum dots,
spintronic devices etc. This effort aimed not only at providing explanations for
experimental observations but also at predicting new effects for future
experiments.\cite{ngo2004trilayer,ngo2007}

The above-mentioned aim of this paper is thus to investigate the combined effects
of frustration and film thickness which are expected to be interesting because of the symmetry
reduction. As said above, the bulk FCC Ising antiferromagnet is fully
frustrated because it  is composed of  tetrahedra whose faces are equilateral triangles.
The antiferromagnetic (AF) interaction on such triangles causes a full frustration.\cite{Diep2005}
The bulk properties of
this material have been largely studied as we will show below.
In this paper,  we shall use the recent high precision technique called
"Wang-Landau" flat histogram Monte Carlo (MC)
simulations to identify the order of the transition.
We also use the Green's function (GF) method for qualitative
comparison.

The paper is organized as follows. Section II is devoted to the
description of the model. We recall there properties of the 3D
counterpart model in order to better appreciate properties of thin
films obtained in this paper.   In section III, we show our results obtained by MC
simulations on the order of the transition. A detailed discussion on the
nature of the phase transition is given.
In the regime of second-order transition, we show in this section the
results on the critical exponents obtained by MC flat histogram technique.
  Section IV is devoted to
a study of the quantum version of the same model by the use of the
GF method.  Concluding remarks are given in section V.

\section{Model and Ground State Analysis}

It is known that the AF interaction between
nearest-neighbor (NN) spins on the FCC lattice causes  a very
strong frustration.  This is due to the fact that the FCC lattice
is composed of tetrahedra each of which has four
equilateral triangles.  It is well-known\cite{Diep2005} that it is
impossible to fully satisfy simultaneously the three AF bond
interactions on each triangle.
 In the case of Ising model, the GS is infinitely degenerate for an infinite system size:
on each tetrahedron two  spins are up and the other two are down.  The FCC system is composed of  edge-sharing tetrahedra.  Therefore, there is an infinite number of ways to construct the infinite crystal. The minimum number of ways of such a construction is a stacking,  in one direction, of uncorrelated AF planes.  The minimum GS degeneracy of a $L^3$ FCC-cell system ($L$ being the number of cells in each directions), is therefore equal to $3\times 2^{2L}$ where the factor 3 is the number of choices of the stacking direction, $2$ the degeneracy of the AF spin configuration of each plane and $2L$ the number of atomic planes in
one direction of the  FCC crystal (the total number of spins is $N=4L^3$).  The GS degeneracy is therefore of the order of $2^{N^{1/3}}$.
Note that at finite temperature, due to the so-called "order by disorder",\cite{Villain,Henley} the spins will choose a  long-range ordering. In the case of AF FCC Ising crystal, this ordering is an alternate stacking of up-spin planes and down-spin planes in one of the three direction.  This has been observed also in the Heisenberg case,\cite{Diep1989fcc} as well as in other frustrated systems.\cite{Diep1998}

The phase transition of the bulk frustrated FCC Ising
antiferromagnet has been found to be  of the first order.\cite{Phani,Polgreen,Styer,Pommier,Beath}
Note that for the Heisenberg model, the transition is also found to be of the first order as in
the Ising case.\cite{Diep1989fcc,Gvoz} Other similar
frustrated antiferromagnets such as the HCP XY and Heisenberg antiferromagnets\cite{Diep1992hcp} and
stacked triangular XY and Heisenberg antiferromagnets\cite{Ngo2008STAXY,Ngo2008STAH}  show
the same behavior.

Let us consider a film of FCC lattice structure with (001)
surfaces.  The Hamiltonian is given by
\begin{equation}
\mathcal H=-\sum_{\left<i,j\right>}J_{i,j}\sigma_i\cdot\sigma_j \label{eqn:hamil1}
\end{equation}
where $\sigma_i$ is the Ising spin at the lattice site
$i$, $\sum_{\left<i,j\right>}$ indicates the sum over the NN spin
pairs  $\sigma_i$ and $\sigma_j$.

In the following, the interaction between two NN on the surface is supposed to be
AF and equal to $J_s$. All other interactions are equal to $J=-1$ for simplicity.
Note that in a previous paper,\cite{Ngo07fccfilm} we have studied the case of the Heisenberg model on the  same
FCC AF film as a function of $J_s$.

\begin{figure}
\centerline{\epsfig{file=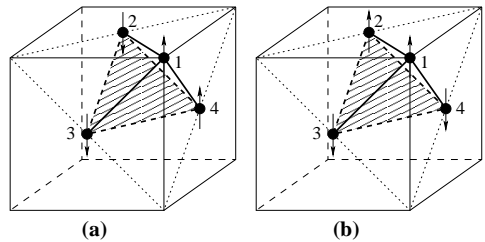,width=2.5in}} \caption{The
ground state spin configuration of the FCC cell at the film
surface: a) ordering of type I for $J_s<-0.5|J|$ b) ordering of type
II for $J_s>-0.5|J|$.} \label{fig:gsstruct}
\end{figure}

For Ising spins, the GS configuration can be determined in a simple way as follows: we calculate the energy of the
surface spin in the two configurations shown in Fig. \ref{fig:gsstruct}  where  the film surface
contains spins 1 and 2 while the beneath layer spins 3 and 4. In the  ordering of type I [Fig. \ref{fig:gsstruct}(a)],
the spins on the surface ($xy$ plane) are antiparallel  and  in the ordering of type II [Fig. \ref{fig:gsstruct}(b)]
they are parallel. Of course, apart from
the overall inversion, for type I there is a degenerate configuration by exchanging the spins 3 and 4.  To see
which configuration is stable, we write the energy of a surface spin for these two configurations
\begin{eqnarray}
E_I&=&-4|J_s|\nonumber\\
E_{II}&=&4|J_s|-4|J| \label{eqn:totEplaq}
\end{eqnarray}
One sees that $E_{I}\leq E_{II}$ when $J_s/J \geq 0.5$.  In the following, we study the case
$J_s=J=-1$ so that the GS configuration is of type I.

\section{Monte Carlo Results}
In this paragraph, we show the results obtained by MC simulations
with the Hamiltonian (\ref{eqn:hamil1}) using the high-precision Wang-Landau flat
histogram technique.\cite{WL1}
Wang and Landau recently proposed a MC algorithm for classical statistical models.
The algorithm uses a random walk in energy space in order to obtained an accurate
estimate for the density of states $\rho(E)$ which is defined as the number of spin
configurations for any given $E$. This method is based on the fact that a flat energy
histogram $H(E)$ is produced if the probability for the transition to a state of energy $E$ is proportional to $\rho(E)^{-1}$.
At the beginning of the simulation, the density of states (DOS) is set equal to one for all possible energies, $\rho(E)=1$.
We begin a random walk in energy space $(E)$ by choosing a site randomly and flipping its spin with a probability
proportional to the inverse of the momentary density of states. In general, if $E$ and $E'$ are the energies before and after a spin is flipped, the transition probability from $E$ to $E'$ is
\begin{equation}
p(E\rightarrow E')=\min\left[\rho(E)/\rho(E'),1\right].
\label{eq:wlprob}
\end{equation}
Each time an energy level $E$ is visited, the DOS is modified by a modification factor $f>0$ whether the spin flipped or not, i.e. $\rho(E)\rightarrow \rho(E)f$.
  At the beginning of the random walk, the modification factor $f$ can be as large as $e^1\simeq 2.7182818$. A histogram $H(E)$ records how often a state of energy $E$ is visited. Each time the energy histogram satisfies a certain "flatness" criterion, $f$ is reduced according to $f\rightarrow \sqrt{f}$ and $H(E)$ is reset to zero for all energies. The reduction process of the modification factor $f$ is repeated several times until a final value $f_{\mathrm{final}}$ which close enough to one. The histogram is considered as flat if
\begin{equation}
H(E)\ge x\%.\langle H(E)\rangle
\label{eq:wlflat}
\end{equation}
for all energies, where $x\%$ is chosen between $70\%$ and $95\%$
and $\langle H(E)\rangle$ is the average histogram.

The thermodynamic quantities\cite{WL1,brown} can be evaluated by
\begin{eqnarray}
\langle E^n\rangle &=&\frac{1}{Z}\sum_E E^n \rho(E)\exp(-E/k_BT)\nonumber\\
C_v&=&\frac{\langle E^2\rangle-\langle E\rangle^2}{k_BT^2}\nonumber\\
\langle M^n\rangle &=&\frac{1}{Z}\sum_E M^n \rho(E)\exp(-E/k_BT)\nonumber\\
\chi&=&\frac{\langle M^2\rangle-\langle M\rangle^2}{k_BT}\nonumber
\end{eqnarray}
where $Z$ is the partition function defined by
\begin{equation}
Z =\sum_E \rho(E)\exp(-E/k_BT)
\label{eq:partfunc}
\end{equation}
The canonical distribution at any temperature can be calculated simply by
\begin{equation}
P(E,T) =\frac{1}{Z}\rho(E)\exp(-E/k_BT)
\label{eq:pe}
\end{equation}

In this work, we consider a energy range of interest\cite{Schulz,Malakis}
$(E_{\min},E_{\max})$. We divide this energy range to $R$ subintervals,
the minimum energy of each subinterval is $E^i_{\min}$ for $i=1,2,...,R$,
and maximum of the subinterval $i$ is $E^i_{\max}=E^{i+1}_{\min}+2\Delta E$,
where $\Delta E$ can be chosen large enough for a smooth boundary between two subintervals. The Wang-Landau
algorithm is used to calculate the relative DOS of each subinterval $(E^i_{\min},E^i_{\max})$ with the
modification factor $f_\mathrm{final}=\exp(10^{-9})$ and flatness criterion $x\%=95\%$.
We reject the suggested spin flip and do not update $\rho(E)$ and the energy histogram $H(E)$ of
the current energy level $E$ if the spin-flip trial would result in an energy outside the energy segment.
The DOS of the whole range is obtained by joining the DOS of each
subinterval $(E^i_{\min}+\Delta E,E^i_{\max}-\Delta E)$.

 The film size used in our present work is $L\times L \times N_z$ where
$L$ is the number of cells in $x$ and $y$ directions, while $N_z$ is that along the $z$ direction (film
thickness).  We use
here $L=30, 40, ..., 150$ and $N_z=$ 2, 4, 8, 12. Periodic boundary
conditions are used in the $xy$ planes.  Our computer program was parallelized and run on a rack of several dozens of 64-bit CPU. $|J|=1$ is taken as unit of energy in the
following.

Before showing the results let us adopt the following notations.
Sublattices 1 and 2 of the first FCC cell belongs to the surface layer,
while sublattices 3 and 4 of the first cell belongs to the second
layer [see Fig. \ref{fig:gsstruct}(a)].
In our simulations,
we used  $N_z$ FCC cells, i.e. $2N_z$ atomic layers.  We used the symmetry
of the two film surfaces.

\subsection{Crossover of the phase transition}

\begin{figure}
\centerline{\epsfig{file=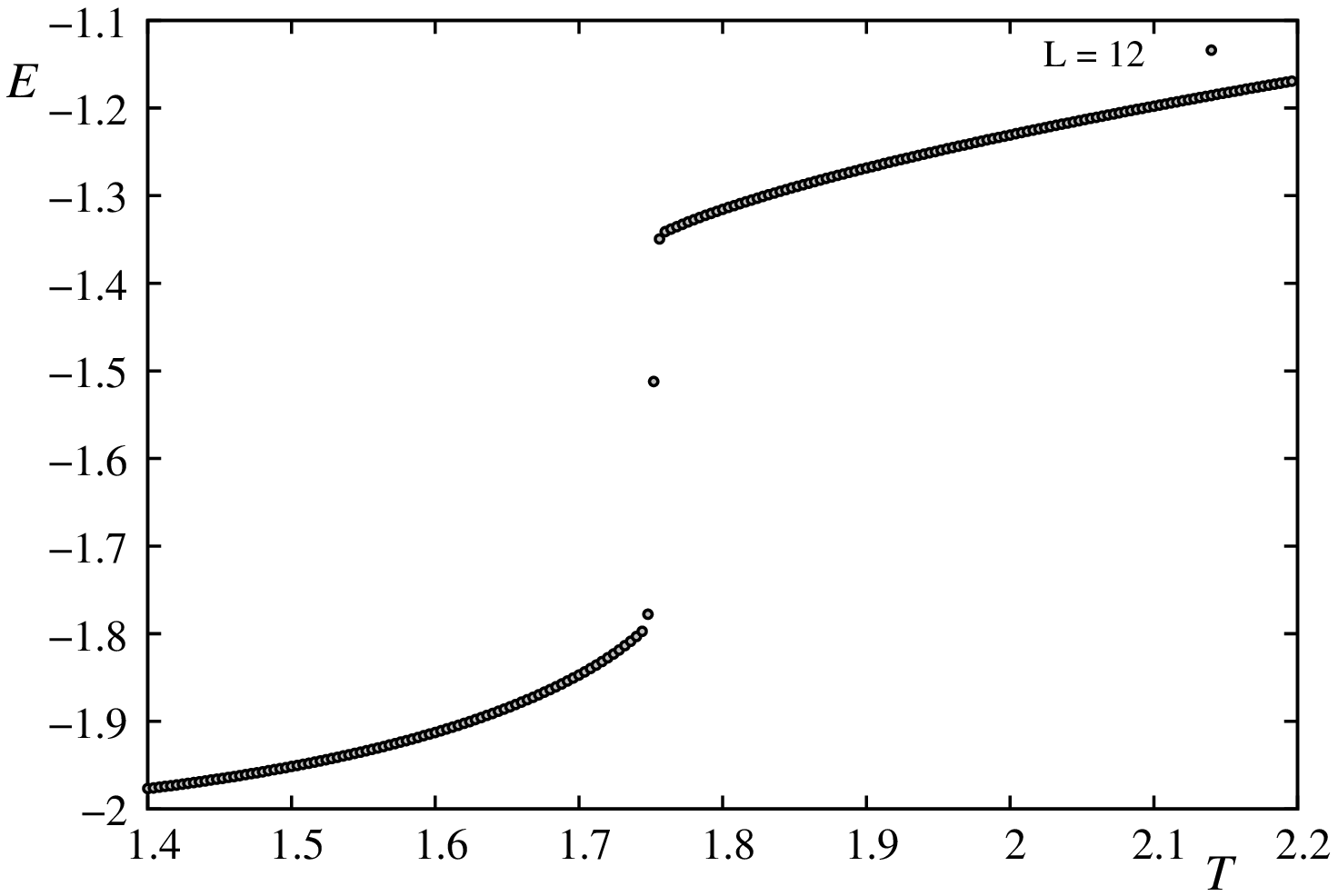,width=2.3in}} \caption{Bulk energy vs $T$
for $L=N_z=12$.}\label{fig:FCC12E}
\centerline{\epsfig{file=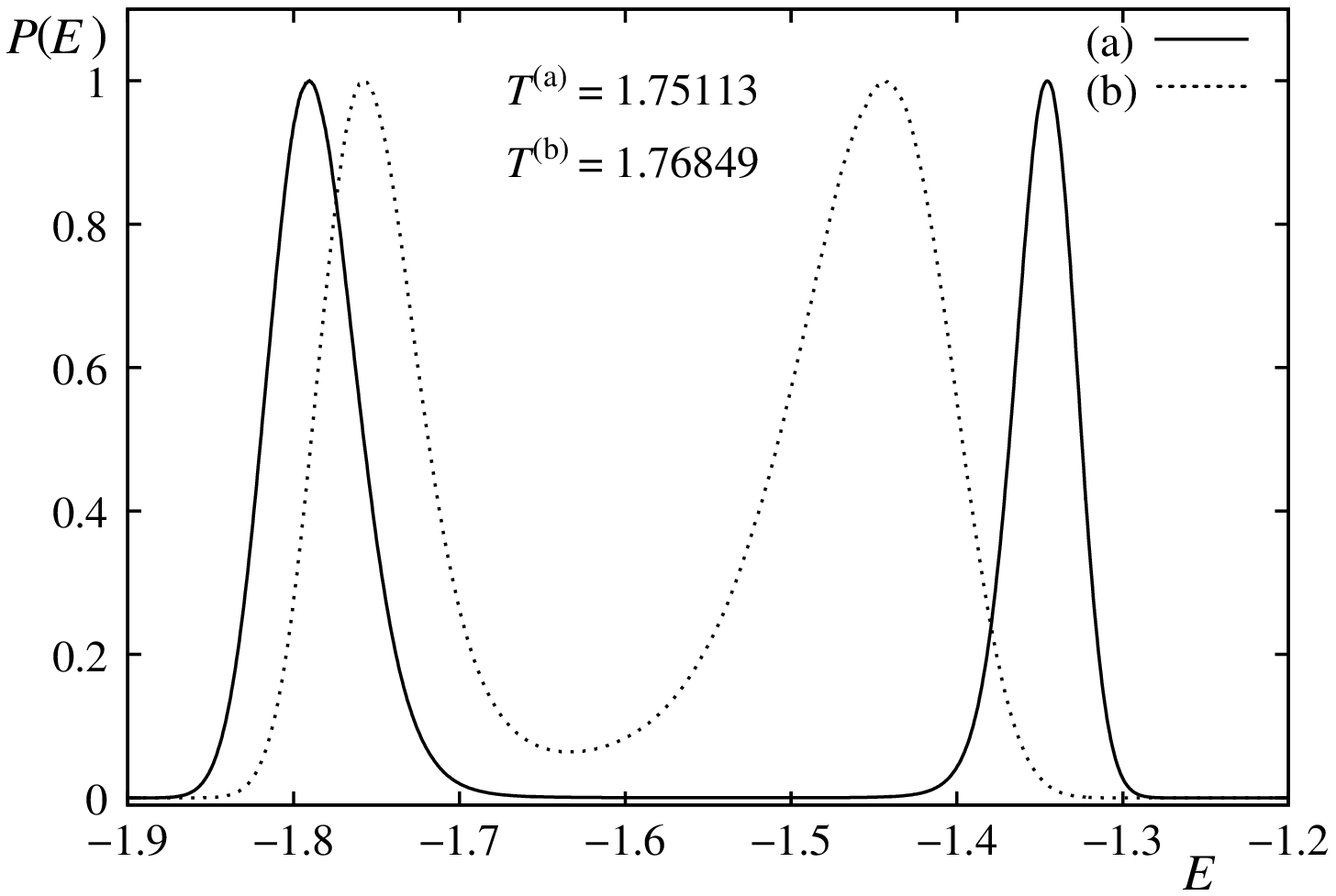,width=2.3in}} \caption{Bulk energy histogram
for $L=N_z=12$ with periodic boundary conditions in all three directions (a) and without PBC in $z$ direction (b). The histogram was taken at
the transition temperature $T_c$ indicated on the figure.} \label{fig:FCC12PE}
\end{figure}

As said earlier, the bulk FCC antiferromagnet with Ising spins shows a very strong first-order
transition. This is seen in MC simulation even with a small lattice size as shown in Fig. \ref{fig:FCC12PE}.

\begin{figure}
\centerline{\epsfig{file=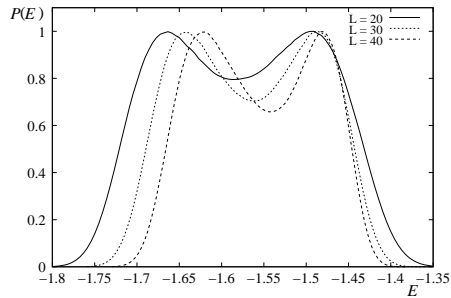,width=2.3in}} \caption{Energy histogram
for $L=20,\ 30,\ 40$ with film thickness $N_z=4$ (8 atomic layers) at $T=1.8218, \ 1.8223, \ 1.8227$, respectively.} \label{fig:Z4PE}
\end{figure}

Our purpose here is to see whether the transition becomes second order when we decrease the film thickness.
As it turns out, the transition remains of first order down to $N_z=4$ as seen by the double-peak energy histogram
displayed in  Fig. \ref{fig:Z4PE}. Note that we do not need to go to larger $L$, the transition is clearly of first order
already at $L=40$.

\begin{figure}
\centerline{\epsfig{file=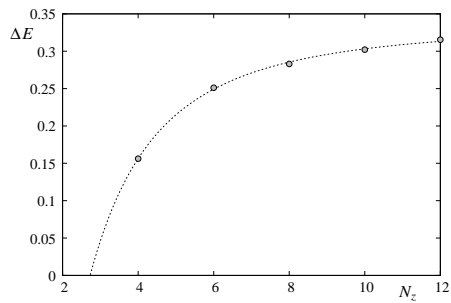,width=2.3in}}
\caption{The latent heat $\Delta E$ as a function of thickness $N_z$.}
\label{fig:DENZ}
\end{figure}

In Fig. \ref{fig:DENZ} we plot the latent heat $\Delta E$ as a function of thickness $N_z$. Data points are well fitted with the following formula

\begin{equation}
\label{eq:defit}
\Delta E=A-\frac{B}{N_z^{d-1}}\left[1+\frac{C}{N_z}\right],
\end{equation}
where $d=3$ is the dimension, $A=0.3370,\ B=3.7068,\ C=-0.8817$. Note that the term $N_z^{d-1}$ corresponds to the surface separating
two domains of ordered and disordered phases at the transition. The second term in the brackets corresponds to a size correction.
As seen in Fig. \ref{fig:DENZ}, the latent heat vanishes at a thickness between 2 and 3.  This is verified by our simulations for $N_z=2$.
For $N_z=2$ we find a  transition with all second-order features: no discontinuity in energy (no double-peak structure)
even when we go up to $L=150$.


Before  showing in the following the results of $N_z=2$, let us discuss on the crossover.
In the case of a film with finite thickness studied here, it
appears that the first-order character is lost for very small $N_z$.
A possible cause for the loss of the first-order transition is from the role of the correlation in
 the film. If a transition is of first order in 3D, i. e. the
correlation length is finite at the transition temperature, then
in thin films the thickness effect may be important: if the
thickness is larger than the correlation length at the transition,
than the first-order transition should remain. On the other hand,
if the thickness is smaller than that correlation length, the
spins then feel an "infinite" correlation length across the film
thickness resulting in a second-order transition.

\subsection{Film with 4 atomic layers ($N_z=2)$}

\begin{figure}
\centerline{\epsfig{file=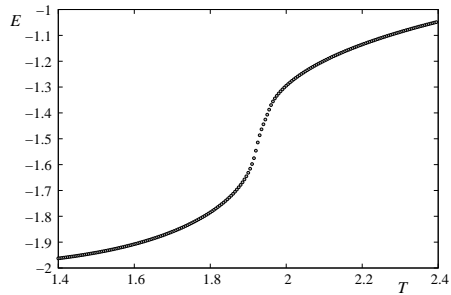,width=2.3in}} \caption{Energy  versus temperature $T$
for $L=120$ with film thickness $N_z=2$.} \label{fig:L120Z2E}
\end{figure}

\begin{figure}
\centerline{\epsfig{file=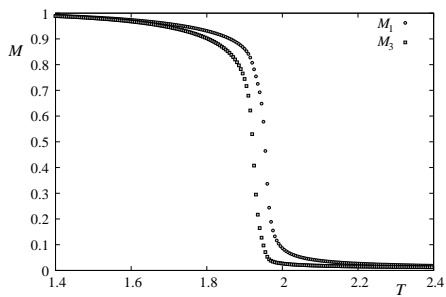,width=2.3in}} \caption{Sublattice magnetization
for $L=120$ with film thickness $N_z=2$.} \label{fig:L120Z2M}
\end{figure}

Let us show in Fig. \ref{fig:L120Z2E} and Fig. \ref{fig:L120Z2M} the energy and the magnetizations of sublattices 1 and 3 of the first two
cells with $L=120$ and $N_z=2$.


It is interesting to note that the surface layer has larger
magnetization than that of the second layer.  This is not
the case for non frustrated films where the surface magnetization
is always smaller than the interior ones because of the effects of
low-lying energy surface-localized magnon
modes.\cite{diep79,diep81}  One explanation can be advanced: due
to the lack of neighbors surface spins are less frustrated than the interior
spins.
As a consequence, the surface spins maintain
their ordering up to a higher temperature.

\begin{figure}
\centerline{\epsfig{file=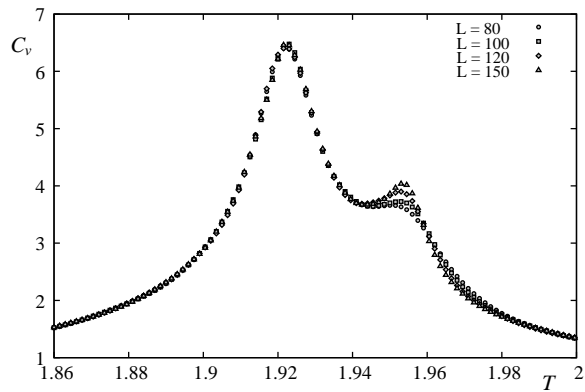,width=3.in}}
\caption{Specific heat are
shown for various sizes $L$ as a function of temperature.} \label{fig:Z2CV}
\end{figure}

\begin{figure}
\centerline{\epsfig{file=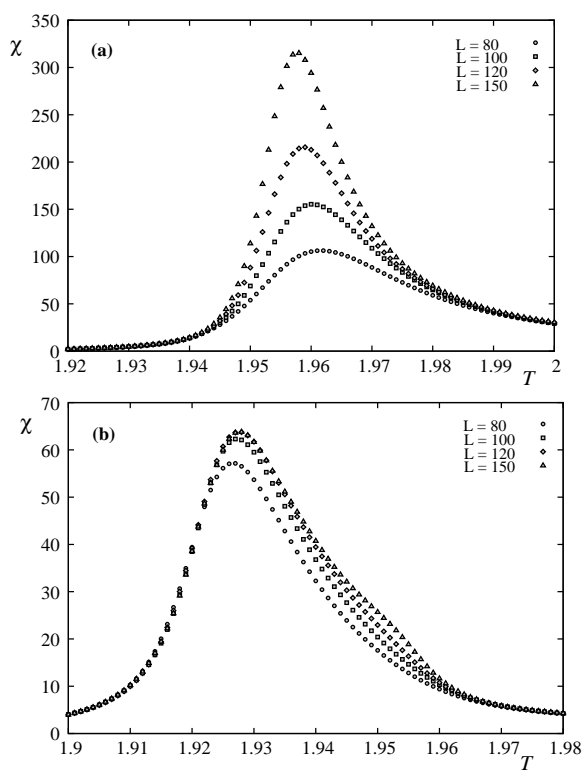,width=3.in}}
\caption{Susceptibilities of sublattice 1 (a) and 3 (b) are
shown for various sizes $L$ as a function of temperature.} \label{fig:Z2X12}
\end{figure}

Let us discuss on finite-size effects in the transitions observed
in Figs. \ref{fig:Z2CV} and \ref{fig:Z2X12}. This is
an important question because it is known that some apparent
transitions are  artifacts of small system sizes.

To confirm further the observed transitions, we have made a study
of finite-size effects on the layer susceptibilities by using the Wang-Landau
technique described above.\cite{WL1}

\begin{figure}[h!]
\centerline{\epsfig{file=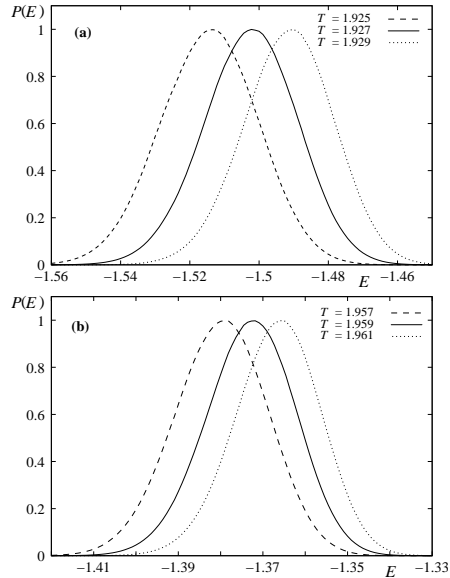,width=2.3in}} \caption{Energy histograms
for $L=120$ with film thickness $N_z=2$ at the two temperatures
(indicated on the figure) corresponding to the  peaks observed in the specific heat.
See text for comment.} \label{fig:L120Z2PE}
\end{figure}

We observe that there are two peaks in the specific heat:   The first peak at $T_1\simeq 1.927$, corresponding to the vanishing of
the sublattice magnetization 3, does not depend on the lattice size while the second peak
at $T_2\simeq 1.959$, corresponding to the vanishing of
the sublattice magnetization 1, does depend on $L$.  Both histograms taken at these temperatures and the near-by ones show
a gaussian form indicating a non first-order transition [see Fig. \ref{fig:L120Z2PE}].

The fact that the peak at $T_1$ does not depend on $L$ suggests two scenarios:

i) the peak does not
correspond to a real transition, since there exist systems where $C_v$ shows a peak but we know that there is no transition just as in the case of 1D Ising chain,

ii) the peak corresponds to a Kosterlitz-Thouless transition.  To confirm this we need to check carefully
several points such as the behavior of the correlation length etc. This is a formidable task which is not
the scope of this work.

Whatever the scenario for the origin of the peak at $T_1$, we know that the interior layers are disordered
between $T_1$ and $T_2$, while the two surface layers are still ordered.   Thus, the transition of the surface layers occurs
while the disordered interior spins  act on them as dynamical random fields.  Unlike
the true 2D random-field Ising model which does not allow a transition at finite temperature,\cite{Imry}
the random fields acting on the surface layer are correlated. This explains why  one has a finite-$T$ transition here.
Note that this situation is known in some exactly  solved models where partial disorder coexists with order
at finite $T$.\cite{diep91honeyc,diep91Kago,Diep-Giacomini}
 However, it is not obvious to
foresee what is the universality class of the transition at $T_2$. The theoretical
argument of Capehart and Fisher\cite{Fisher} does not apply in the present situation because
one does not have a single transition here, unlike the case of simple cubic ferromagnetic films
studied before.\cite{Pham} So, we wish to calculate
the critical exponents associated with the transition at $T_2$.

\begin{figure}
\centerline{\epsfig{file=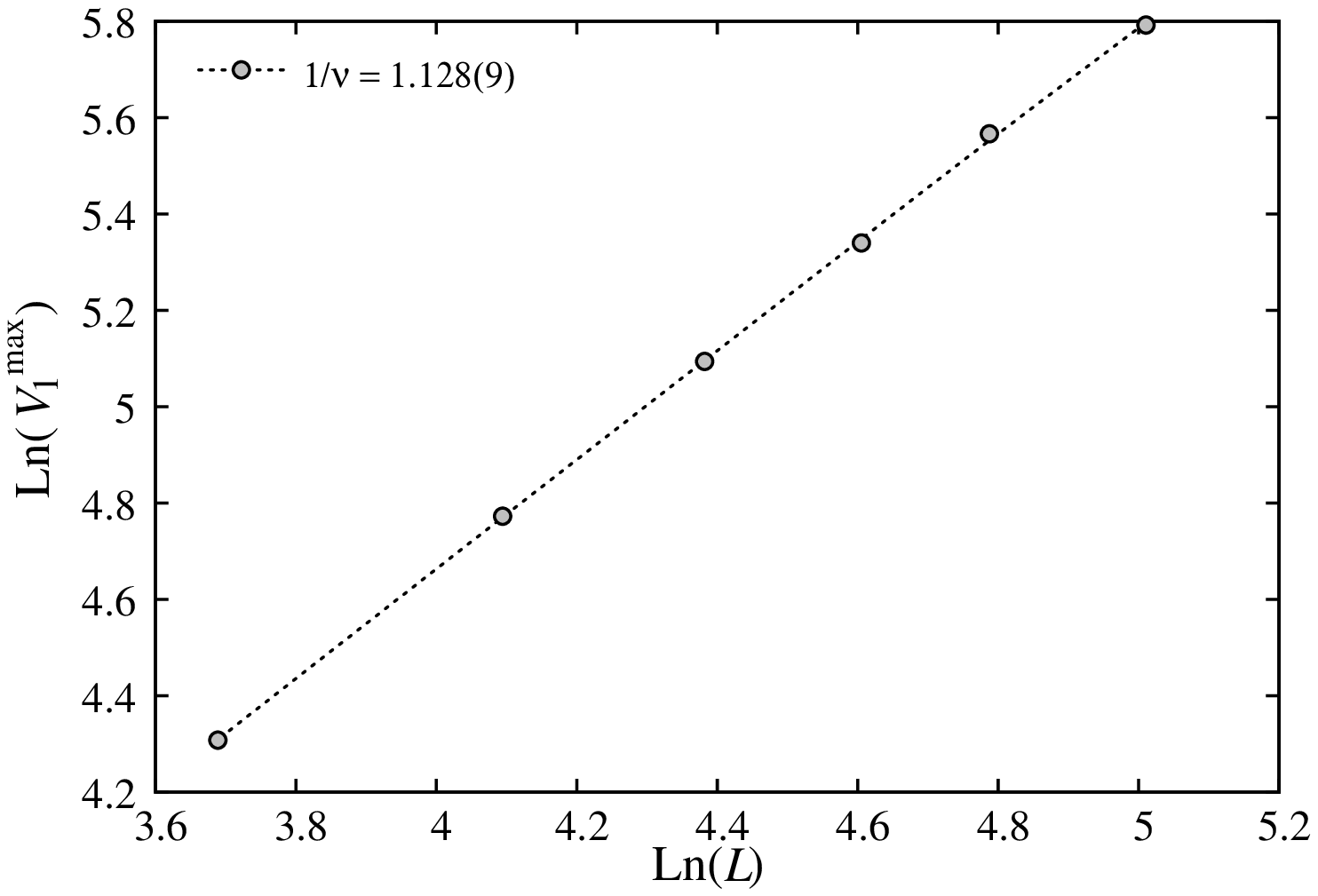,width=2.8in}}
\caption{The maximum value of $V_1=\left<E\right>-\left<ME\right>/\left<M\right>$ versus
$L$ in the $\ln-\ln$ scale. The slope of this straight line
gives $1/\nu$.}
\label{fig:Z2NU}
\end{figure}

\begin{figure}
\centerline{\epsfig{file=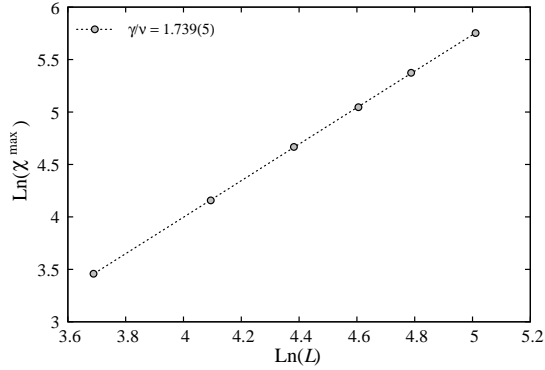,width=2.8in}} \caption{Maximum
sublattice susceptibility  $\chi^{\max}$ versus $L$ in the
$\ln-\ln$ scale. The slope of this straight line
gives $\gamma/\nu$.} \label{fig:Z2GAM}
\end{figure}

 The  exponent $\nu$ can be obtained
as follows. We calculate as a function of $T$ the magnetization
derivative with respect to $\beta=(k_BT)^{-1}$: $V_1=\left<(\ln
M)'\right>=\left<E\right>-\left<ME\right>/\left<M\right>$ where
$E$ is the system energy and $M$ the sublattice order parameter.
We identify the maximum of $V_1$ for each size $L$. From the
finite-size scaling we know that $V_1^{\max}$ is proportional to
$L^{1/\nu}$.\cite{Ferrenberg3}  We show in  Fig. \ref{fig:Z2NU} the
maximum of $V_1$ versus $\ln L$ for the
first layer.  We find $\nu=0.887\pm
0.009$.  Now, using the scaling law $\chi^{\max} \propto L^{\gamma/\nu}$, we
plot $\ln \chi^{\max}$  versus $\ln L$ in Fig. \ref{fig:Z2GAM}.
The ratio of the critical exponents $\gamma/\nu$ is obtained by
the slope of the straight line connecting the data points of each
layer. From the value of $\nu$ we obtain
 $\gamma=1.542\pm 0.005$.
 These values do not correspond neither to 2D nor 3D Ising
models ($\gamma_{2D}=1.75$, $\nu_{2D}=1$, $\gamma_{3D}=1.241$,
$\nu_{3D}=0.63$).  We note however that, if we think of the weak
universality where only ratios of critical exponents are concerned,\cite{Suzuki} then the ratios of these exponents
$1/\nu=1.128$ and $\gamma/\nu=1.739$  are not far from the 2D ones which are 1 and 1.75, respectively.

\section{Green's Function Method}
We consider here the same FCC system but with quantum Heisenberg spins.  To compare the results
with the Ising case, we add in the Hamiltonian an Ising-like
 anisotropy interaction term. In addition, this term  avoids the absence of long-range order of isotropic
non Ising spin model at finite temperature ($T$) when the film
thickness is very small, i.e. quasi two-dimensional
system.\cite{Mermin} The Hamiltonian is given by
\begin{equation}
\mathcal H=-\sum_{\left<i,j\right>}J_{i,j}\mathbf S_i\cdot\mathbf
S_j -\sum_{\left< i,j\right>} I_{i,j}S^z_iS^z_j  \label{eqn:hamil2}
\end{equation}
where $\mathbf S_i$ is the Heisenberg spin at the lattice site
$i$, $\sum_{\left<i,j\right>}$ indicates the sum over the NN spin
pairs  $\mathbf S_i$ and $\mathbf S_j$.  $J_{i,j}$ and $I_{i,j}$ are antiferromagnetic (negative).
Note that in the laboratory coordinates, the up spins have $S^z\geq 0$ while the down spins have
$S^z\leq 0$.

We can rewrite the Hamiltonian (\ref{eqn:hamil2}) in the relative
local spin coordinates as
\begin{eqnarray}
\mathcal H &=& - \sum_{<i,j>}
J_{i,j}\Bigg\{\frac{1}{4}\left(\cos\theta_{ij} -1\right)
\left(S^+_iS^+_j +S^-_iS^-_j\right)\nonumber\\
&+& \frac{1}{4}\left(\cos\theta_{ij} +1\right) \left(S^+_iS^-_j
+S^-_iS^+_j\right)\nonumber\\
&+&\frac{1}{2}\sin\theta_{ij}\left(S^+_i +S^-_i\right)S^z_j
-\frac{1}{2}\sin\theta_{ij}S^z_i\left(S^+_j
+S^-_j\right)\nonumber\\
&+&\cos\theta_{ij}S^z_iS^z_j\Bigg\}- \sum_{<i,j>}I_{i,j}\cos\theta_{ij}S^z_iS^z_j
\label{eq:HGH2}
\end{eqnarray}
where $\theta_{ij}$ is the angle between two NN
spins.  Note that in the above expression, we have transformed all $S^z \geq 0$, the relative spin orientation of each spin pair is now expressed by $\theta_{ij}$.  In a collinear spin configuration such as those shown in Fig. \ref{fig:gsstruct},  $\cos\theta_{ij}=-1$ and $1$ for antiparallel and parallel pairs, respectively, while $\sin\theta_{ij}=0$.  In non collinear structures, the calculation is more complicated. The general GF method for non collinear spin configuration
has been proposed elsewhere.\cite{Rocco,Santamaria}
In the present study, one has a
collinear spin configuration shown in Fig. \ref{fig:gsstruct} because of the Ising-like anisotropy.
We define two double-time
GF by
\begin{eqnarray}
G_{ij}(t,t')&=& \ll S^{+}_i(t) ; S^-_j(t')\gg, \\
F_{mj}(t,t')&=&\ll S^{-}_m(t) ; S^-_j(t')\gg.
\end{eqnarray}
where $i$ and $j$ belong to the up-spin sublattice, $m$ to the down-spin one.
In the case of thin films,
the reader is referred to Refs. \onlinecite{diep79,diep81,Ngo07fccfilm} for
a general formulation.
We describe here only the
main steps: we first
write the equations of motion for $G_{ij}(t,t')$ and $F_{mj}(t,t')$ and
we next neglect higher-order correlations by using the Tyablikov
decoupling scheme\cite{Tyablikov} which is known to be valid for
exchange terms,\cite{fro}   and then we introduce the following Fourier
transforms in the $xy$ plane

\begin{eqnarray}
G_{i, j}\left( t, t'\right) &=& \frac {1}{\Delta}\int\int d\mathbf
k_{xy}\frac{1}{2\pi}\int^{+\infty}_{-\infty}d\omega e^{-i\omega
\left(t-t'\right)}.\nonumber\\
&&\hspace{0.7cm}g_{n,n'}\left(\omega , \mathbf k_{xy}\right)
e^{i\mathbf k_{xy}\cdot \left(\mathbf R_i-\mathbf
R_j\right)},\label{eq:HGFourG}\\
F_{m, j}\left( t, t'\right) &=& \frac {1}{\Delta}\int\int d\mathbf
k_{xy}\frac{1}{2\pi}\int^{+\infty}_{-\infty}d\omega e^{-i\omega
\left(t-t'\right)}.\nonumber\\
&&\hspace{0.7cm}f_{n,n'}\left(\omega , \mathbf k_{xy}\right)
e^{i\mathbf k_{xy}\cdot \left(\mathbf R_m-\mathbf
R_j\right)},\label{eq:HGFourF}
\end{eqnarray}
where $\omega$ is the spin-wave frequency, $\mathbf k_{xy}$
denotes the wave-vector parallel to $xy$ planes, $\mathbf R_i$ is
the position of the spin at the site $i$, $n$ and $n'$ are
respectively the indices of the layers where the sites $i$ (or $m$) and $j$
belong to. One has $n,n'=1,2,...,2N_z$. The integral over $\mathbf k_{xy}$ is performed in the
first Brillouin zone in the $xy$
reciprocal plane whose surface is $\Delta$.  Finally, one obtains for all layers
the following matrix equation
\begin{equation}
\mathbf M \left( \omega \right) \mathbf g = \mathbf u,
\label{eq:HGMatrix}
\end{equation}
Note that though $n$ runs from 1 to $2N_z$, the matrix $\mathbf M$ has the dimension of $4N_z\times 4N_z$ because
for each $n$ there are two functions $g(n,n')$ and $f(n,n')$. In the above equation, $\mathbf g$ and $\mathbf u$ are
the  column matrices of dimension $4N_z$ which are defined as follows
\begin{equation}
\mathbf g = \left(%
\begin{array}{c}
  g_{1,n'} \\
  f_{1,n'} \\
  \vdots \\
  g_{2N_z,n'} \\
  f_{2N_z,n'} \\
\end{array}%
\right) , \mathbf u =\left(%
\begin{array}{c}
  2 \left< S^z_1\right>\delta_{1,n'} \\
  0 \\
  \vdots \\
  2 \left< S^z_{2N_z}\right>\delta_{2N_z,n'} \\
  0 \\
\end{array}%
\right) , \label{eq:HGMatrixgu}
\end{equation}
and
\begin{center}
\begin{equation}
\mathbf M\left(\omega\right) = \left(%
\begin{array}{ccccccc}
  A^+_1    & B_1    &  D^+_1 &  D^-_1 & \cdots & \cdots& \cdots\\
  -B_1     & A^-_1  & -D^-_1 & -D^+_1  & \cdots & \cdots & \cdots\\
   C^+_2  & C^-_2 & A^+_2 & B_2 &D^+_2 & D^-_2& \cdots\\
   - C^-_2  & -C^+_2 &  -B_2& A^-_2 & -D^-_2 & -D^+_2&\cdots\\
  \cdots &\cdots &\cdots &\cdots &\cdots &\cdots &\cdots \\
  \cdots &\cdots & \cdots  & C^+_{N_z}   & C^-_{N_z}   & A^+_{N_z}      & B_{N_z}\\
  \cdots &\cdots   & \cdots     & -C^-_{N_z}  & -C^+_{N_z}  & -B_{N_z}       & A^-_{N_z}\\
\end{array}%
\right) \label{eq:HGMatrixM}
\end{equation}
\end{center}
where for the spin configuration of type I [Fig. \ref{fig:gsstruct} (a)]  one has
\begin{eqnarray}
A_n^\pm &=&  \omega \pm\Bigg[2J_n \left< S^z_n\right>Z + 8I_{n}\left< S^z_n\right>\nonumber \Bigg]\\
B_n &=& -2J_{n}\left<S^z_n\right>\left(Z\gamma\right)\\
C_n^+ &=& +4J_{n,n-1}\left<S^z_n\right>\cos\frac{k_ya}{2}\\
C_n^- &=& -4J_{n,n-1}\left<S^z_n\right>\cos\frac{k_xa}{2}\\
D_n^+ &=& +4J_{n,n+1}\left<S^z_n\right>\cos\frac{k_ya}{2}\\
D_n^- &=& -4J_{n,n+1}\left<S^z_n\right>\cos\frac{k_xa}{2}
\end{eqnarray}
in which, $Z=4$ is the number of in-plane NN, and
$$\gamma =\frac{1}{Z}\left[ 4\cos \left( \frac{k_x a}{2} \right)\cos \left( \frac{k_y a}{2} \right)\right].$$

Here, for compactness we have used the following notations:

i) $J_n$ and $I_n$ are the in-plane interactions. In the present
model $J_n$ is equal to $J=-1$. All $I_n$ are set to be $I(<0)$.

ii) $ J_{n,n\pm 1}$ are the interactions between a spin in the
$n$-th layer and its neighbor in the $(n\pm 1)$-th layer. Here, we take $ J_{n,n\pm 1}=-1$. Of
course, $ J_{n,n-1}=0$ if $n=1$, $ J_{n,n+1}=0$ if $n=2N_z$.

Now, solving det$(\mathbf M)\equiv|\mathbf M|=0$, we obtain the spin-wave spectrum
$\omega$ of the present system.  For each $\mathbf k_{xy}$ there are $4N_z$ eigenvalues $\omega$,
two by two with opposite signs because of the AF symmetry.
The solution for the GF $g_{n,n}$ is given by
\begin{equation}
g_{n,n} = \frac{\left|\mathbf M\right|_n}{\left|\mathbf M\right|},
\end{equation}
with $\left|\mathbf M\right|_n$ is the determinant made by
replacing the $n$-th column of $\left|\mathbf M\right|$ by
$\mathbf u$ in (\ref{eq:HGMatrixgu}). Writing now
\begin{equation}
\left|\mathbf M\right| = \prod_i \left(\omega -
\omega_i\left(\mathbf k_{xy}\right)\right),
\end{equation}
one sees that $\omega_i\left(\mathbf k_{xy}\right) ,\ i = 1,\cdots
,\ 4N_z$, are poles of the GF $g_{n,n}$.
Now, we can express $g_{n,n}$ as
\begin{equation}
g_{n, n} = \sum_i\frac {f_n\left(\omega_i\left(\mathbf
k_{xy}\right)\right)}{\left( \omega - \omega_i\left(\mathbf
k_{xy}\right)\right)}, \label{eq:HGGnn}
\end{equation}
where $f_n\left(\omega_i\left(\mathbf k_{xy}\right)\right)$ is
\begin{equation}
f_n\left(\omega_i\left(\mathbf k_{xy}\right)\right) = \frac{\left|
\mathbf M\right|_n \left(\omega_i\left(\mathbf
k_{xy}\right)\right)}{\prod_{j\neq i}\left(\omega_j\left(\mathbf
k_{xy}\right)-\omega_i\left(\mathbf k_{xy}\right)\right)}.
\end{equation}

Next, using the spectral theorem which relates the correlation
function \(\langle S^-_i S^+_j\rangle \) to the GF,\cite{zu} one has
\begin{eqnarray}
\left< S^-_i S^+_j\right> &=& \lim_{\varepsilon\rightarrow 0}
\frac{1}{\Delta}\int\int d\mathbf k_{xy}
\int^{+\infty}_{-\infty}\frac{i}{2\pi}\big( g_{n, n'}\left(\omega
+ i\varepsilon\right)\nonumber\\
&-& g_{n, n'}\left(\omega - i\varepsilon\right)\big)
\cdot\frac{d\omega}{e^{\beta\omega} - 1}e^{i\mathbf
k_{xy}\cdot\left(\mathbf R_i -\mathbf R_j\right)},
\end{eqnarray}
where $\epsilon$ is an  infinitesimal positive constant and
$\beta=1/k_BT$, $k_B$ being the Boltzmann constant.

Using the GF presented above, we can calculate
self-consistently various physical quantities as functions of
temperature $T$.    Large values of Ising-like interaction $I$ will enhance the ordering. On the contrary,
for $I\rightarrow 0$ the transition temperature will go to zero according to the Mermin-Wagner theorem.\cite{Mermin} This is seen in the following.  For numerical integretaion,
we will use $80^2$ points in the
first Brillouin zone.

\begin{figure}
\centerline{\epsfig{file=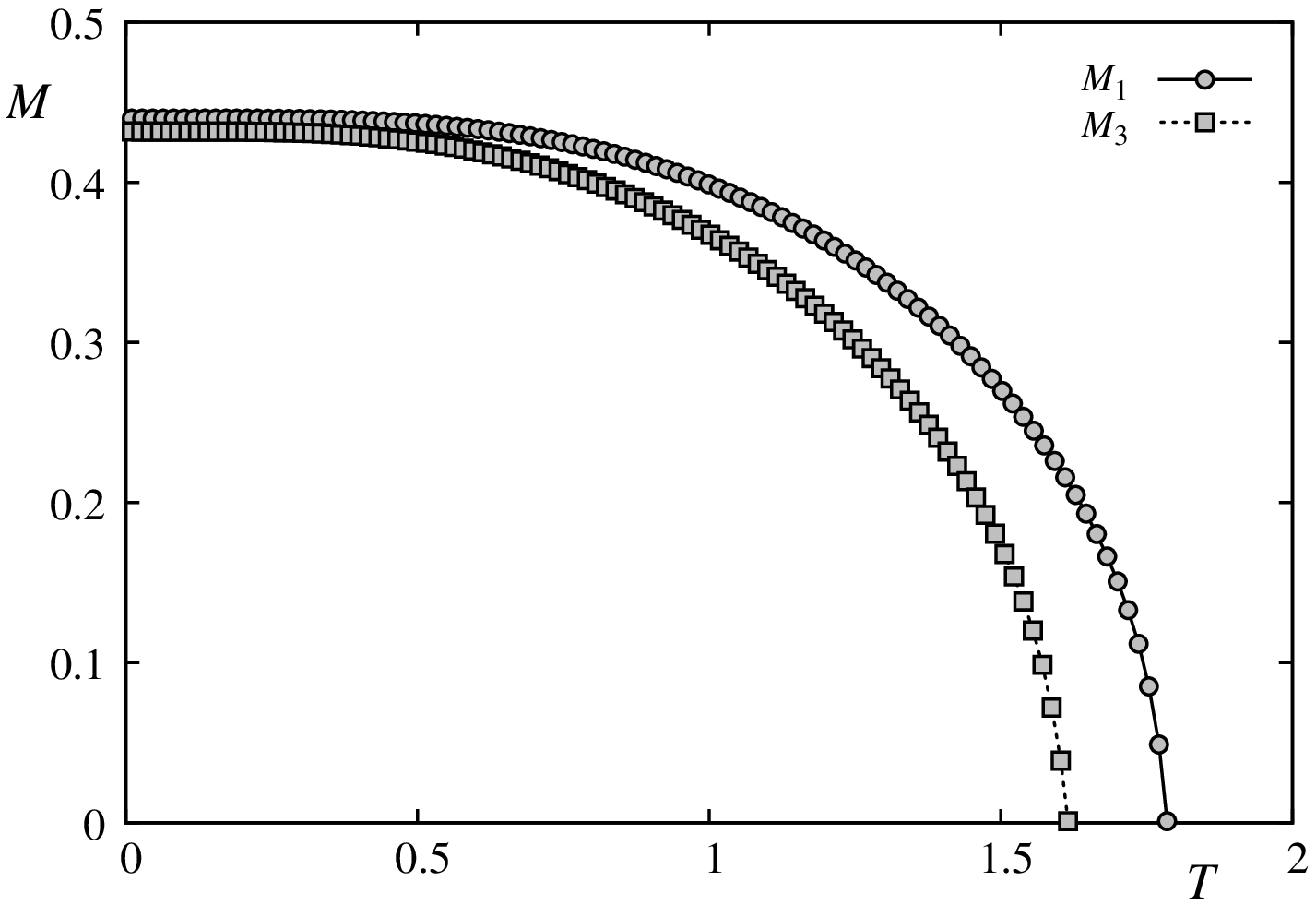,width=2.8in}}
\centerline{\epsfig{file=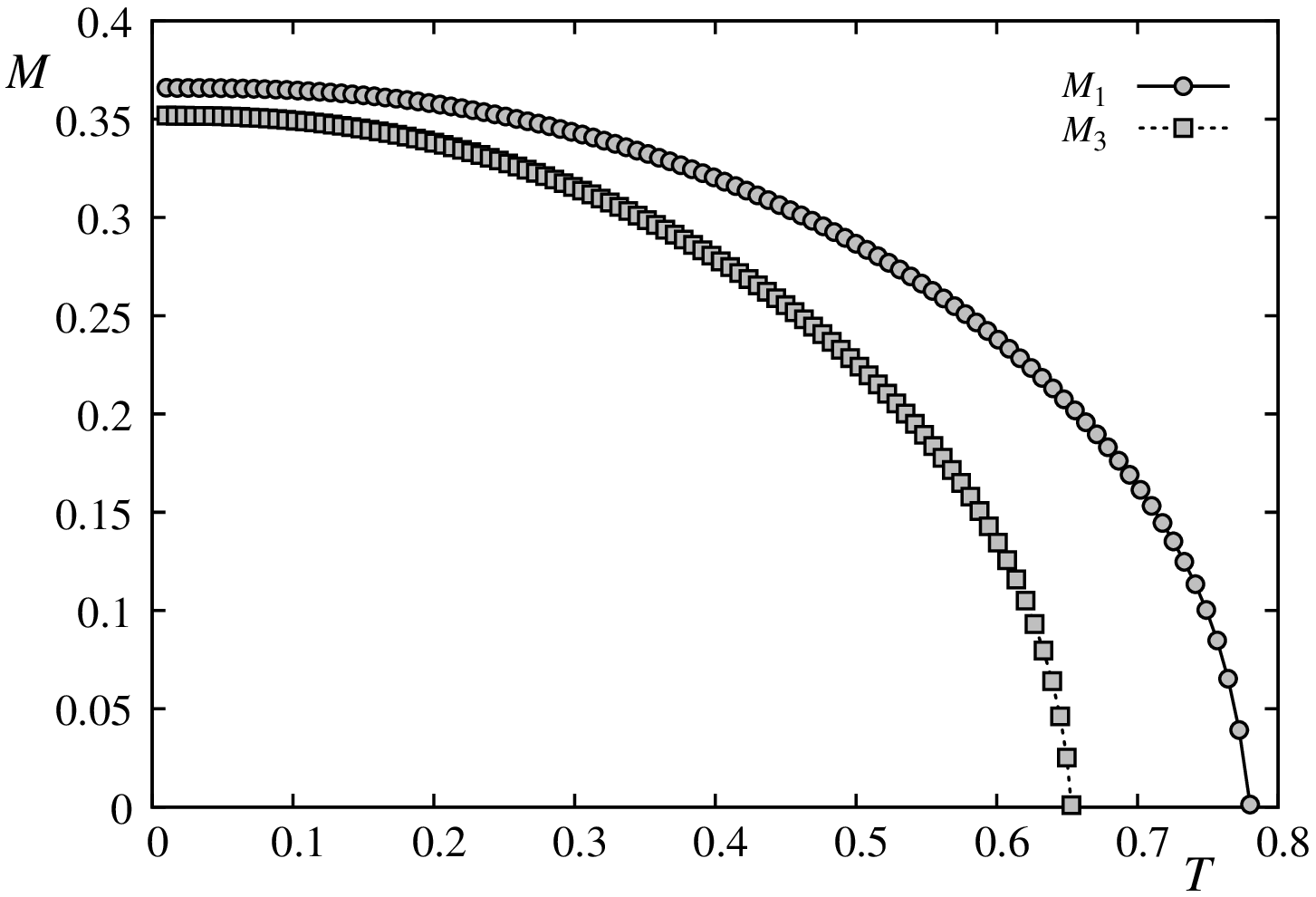,width=2.8in}}
\caption{Magnetization of sublattices 1 (surface) and 3 (second layer) versus temperature for $N_z=2$ and $I=-0.25$ (upper) and $I=-0.01$ (lower).}
\label{fig:GFD4}
\end{figure}

Figure  \ref{fig:GFD4} shows the sublattice magnetizations of the
first two layers for $N_z=2$ with $I=-0.25$ and $I=-0.01$ (upper and lower figures, respectively)). As seen, the surface sublattice magnetization is larger than the
sublattice magnetization of the
second layer for $N_z=2$ in qualitative agreement with the MC results shown in
Fig. \ref{fig:L120Z2M}, in spite of the fact that due to a finite-size effect, there is a queue of the sublattice magnetization above
the transition temperatures for MC results. Note that the AF coupling gives rise to a zero-point spin
contraction  at $T=0$ which is different for the surface spins and the second-layer spins.

\begin{figure}
\centerline{\epsfig{file=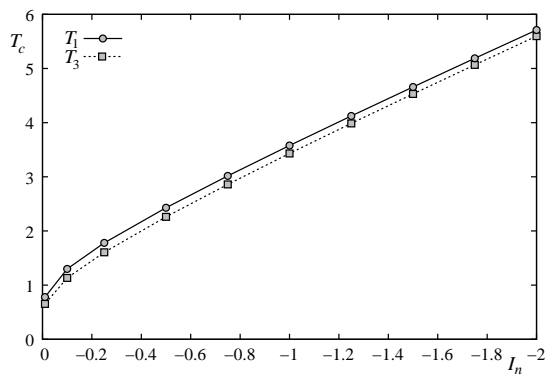,width=2.8in}}
\caption{Phase diagram obtained by the GF method.  $T_1$ and $T_3$ are transition temperatures of the sublattice 1 of the surface and of the sublattice 3 of the second layer.  The transition temperatures should go to zero as $I\rightarrow 0$ (see text).}
\label{fig:GFD5}
\end{figure}

We show in Fig. \ref{fig:GFD5} the phase diagram in the space ($I,T)$ where $T_1$ and $T_3$ are transition temperatures of the surface sublattice 1 and the sublattice 3 of the second layer.

\section{Concluding Remarks}

We have shown in this paper the crossover of the phase transition from first to second order in
the frustrated Ising FCC AF film.  This crossover occurs when the film thickness $N_z=2$
is smaller than a value between 2 and 4 FCC lattice cells.  These results are obtained with the highly performing Wang-Landau flat
histogram technique which allows to determine the first-order transition with efficiency.

For $N_z=2$, we found that in a range of temperature the surface spins stay ordered  while interior spins are disordered. We interpret this as an effect of the frustration reduction: due to the lack of neighbors, the surface spins are less  frustrated
than the interior spins. As a consequence, interior spins are disordered at a lower temperature.  This has been verified by the Green's function calculation.

The second-order transition for $N_z=2$ is governed by the surface disordering and is characterized by  critical exponents whose values are deviated from those of the 2D Ising universality class. We believe that this deviation  results from the effect of the disordered interior spins which act as "correlated" random fields on the surface spins.  We do not know if the critical exponents found here belong to a new universality class or they are just "effective critical exponents" which one could scale in some way or another to bring into the 2D Ising universality class.  Anyway, these exponents seem to obey a weak universality.  An answer to this question is desirable.

One of us (VTN) thanks the University of Cergy-Pontoise for
 a financial support and hospitality during the final stage of this work.

{}


\begin{thebibliography}{}


\bibitem{Pham} X. T. Pham Phu, V. Thanh Ngo and H. T. Diep, Surf. Science {\bf 603}, 109 (2009).

\bibitem{Fisher}  T. W. Capehart and M. E. Fisher,
 Phys. Rev.  B {\bf 13}, 5021 (1976).

\bibitem{diep91honeyc} H. T. Diep, M. Debauche and H. Giacomini, Phys. Rev. B (rapid communication) {\bf 43}, 8759 (1991).

\bibitem{diep91Kago} M. Debauche, H. T. Diep, H. Giacomini and P. Azaria, Phys. Rev. B {\bf 44}, 2369 (1991).

 \bibitem{Diep2005} See reviews on theories and experiments given in
{\it Frustrated Spin Systems}, ed. H. T. Diep, World Scientific
(2005).


\bibitem{zangwill} A. Zangwill, {\it Physics at Surfaces},
Cambridge University Press (1988).

\bibitem{bland-heinrich} {\it Ultrathin Magnetic Structures}, vol. I and II,
J.A.C. Bland and B. Heinrich (editors), Springer-Verlag (1994).


\bibitem{Diehl}H.W. Diehl, in {\it Phase Transitions and
Critical Phenomena}, ed. by C. Domb, J.L. Lebowitz (Academic,
London, 1986) vol. 10, H.W. Diehl, Int. J. Mod. Phys. B {\bf 11},
3503 (1997).

\bibitem{Baibich} M. N. Baibich, J. M. Broto, A. Fert, F. Nguyen
Van Dau, F. Petroff, P. Etienne, G. Creuzet, A. Friederich and J.
Chazelas, Phys. Rev. Lett. {\bf 61}, 2472 (1988).

\bibitem{Grunberg} P. Grunberg, R. Schreiber, Y. Pang, M. B. Brodsky and H. Sowers,
Phys. Rev. Lett. {\bf 57}, 2442 (1986); G. Binash, P. grunberg, F.
Saurenbach and W. Zinn, Phys. Rev. B {\bf 39}, 4828 (1989).

\bibitem{Fert} A. Barth\'el\'emy et al, J. Mag.  Mag. Mater. {\bf
242-245}, 68 (2002).

\bibitem{review} See review by E. Y. Tsymbal and D. G. Pettifor, {\it Solid
State Physics} (Academic Press, San Diego), Vol. 56, pp. 113-237
(2001).

\bibitem{ngo2004trilayer} See V. Thanh Ngo, H. Viet Nguyen, H. T. Diep and V. Lien Nguyen,
 Phys. Rev. B. {\bf 69}, 134429 (2004) and references on magnetic multilayers cited therein.

\bibitem{ngo2007} See V. Thanh Ngo and H. T. Diep,
 Phys. Rev. B. {\bf 75}, 035412 (2007) and references on surface effects cited therein.




\bibitem{Villain} J. Villain, R. Bidaux, J. P. Carton, and R. Conte, J. Phys. (Paris) {\bf
41}, 1263 (1980).

\bibitem{Henley} C. L. Henley,  J. Appl. Phys. {\bf
61}, 3962 (1987).

\bibitem{Diep1989fcc} H. T. Diep and H. Kawamura, Phys. Rev. B {\bf
40}, 7019 (1989).

\bibitem{Diep1998} C. Pinettes and H. T. Diep, J. Appl. Phys. {\bf
83}, 6317 (1998).

\bibitem{Phani} J. L. Lebowitz and M. H. Kalos, Phys. Rev. B {\bf
21}, 4027 (1980).

\bibitem{Polgreen} T. L. Polgreen, Phys. Rev. B {\bf
29}, 1468 (1984).

\bibitem{Styer} D. F. Styer, Phys. Rev. B {\bf
32}, 393 (1985).

\bibitem{Pommier} J. Pommier, H. T. Diep, A. Ghazali and P. Lallemand,
J. Appl. Phys. {\bf 63}, 3036 (1988).

\bibitem{Beath} A. D. Beath and D. H. Ryan, Phys. Rev. B {\bf
73}, 174416 (2006).

\bibitem{Gvoz} M. V. Gvozdikova and M. E. Zhitomirsky, JETP Lett. {\bf 81}, 236
(2005).

\bibitem{Diep1992hcp} H. T. Diep, Phys. Rev. B {\bf
45}, 2863 (1992), and references therein.

\bibitem{Ngo2008STAXY}  V. Thanh Ngo and H. T. Diep, J. Appl. Phys. {\bf 103}, 07C712 (2008).

\bibitem{Ngo2008STAH}  V. Thanh Ngo and H. T. Diep, Phys. Rev. E,   {\bf 78}, 031119 (2008).


\bibitem{Ngo07fccfilm}  V. Thanh Ngo and H. T. Diep, J. Phys.: Condens. Matter (2007)



\bibitem{WL1} F. Wang and D. P. Landau, Phys. Rev. Lett. {\bf
86}, 2050 (2001); Phys. Rev. E {\bf 64}, 056101 (2001).

\bibitem{brown} G. Brown and T.C. Schulhess, J. Appl. Phys.
{\bf 97}, 10E303 (2005).
\bibitem{Schulz}B. J. Schulz, K. Binder, M. M\"{u}ller, and D. P.
Landau, Phys. Rev. E {\bf 67}, 067102 (2003).
\bibitem{Malakis} A. Malakis, S. S. Martinos, I. A. Hadjiagapiou, N. G.
Fytas, and P. Kalozoumis, Phys. Rev. E {\bf 72}, 066120 (2005).



\bibitem{diep79}
Diep-The-Hung, J.C. S. Levy and O. Nagai, Phys. Stat. Solidi (b), {\bf
93}, 351 (1979).

\bibitem{diep81}Diep-The-Hung,
Phys. Stat. Solidi (b), {\bf 103}, 809 (1981).

\bibitem{Imry} Yoseph Imry and Shang-Keng Ma, Phys. Rev. Lett. {\bf 35}, 1399 (1975).

\bibitem{Diep-Giacomini} H. T. Diep and H. Giacomini, see chapter {\it Frustration - Exactly Solved Frustrated Models} in Ref. \onlinecite{Diep2005}.



\bibitem{Ferrenberg3}
A. M. Ferrenberg and D. P. Landau, Phys. Rev. B {\bf 44}, 5081
(1991).

\bibitem{Suzuki} Masuo Suzuki, Prog. Theor. Phys. {\bf 51}, 1992 (1974).









\bibitem{Mermin}  N. D. Mermin and H. Wagner, Phys. Rev. Lett. {\bf 17},
1133 (1966).

\bibitem{Rocco} See for example
R. Quartu and H.T. Diep, Phys. Rev. B {\bf 55}, 2975 (1997).

\bibitem{Santamaria}
C. Santamaria, R. Quartu and H. T. Diep,  J. Appl. Physics  {\bf 84}, 1953 (1998).

\bibitem{Tyablikov} N. N. Bogolyubov and S. V. Tyablikov, Doklady Akad.
Nauk S.S.S.R. {\bf 126}, 53 (1959) [translation: Soviet
Phys.-Doklady {\bf 4} 604 (1959)].

\bibitem{fro} P. Fr\" {o}brich, P. J. Jensen and P. J. Kuntz,
Eur. Phys. J B {\bf 13}, 477 (2000) and references therein.

\bibitem{zu} D. N. Zubarev, Usp. Fiz. Nauk {\bf 187}, 71
(1960)[translation: Soviet Phys.-Uspekhi {\bf 3} 320 (1960)].





\end{thebibliography}
\end{document}